# On the Validity of Covariate Adjustment for Estimating Causal Effects

**Ilya Shpitser**
Department of Epidemiology
Harvard School of Public Health
ishpitse@hsph.harvard.edu

**Tyler VanderWeele**
Department of Epidemiology
Department of Biostatistics
Harvard School of Public Health
tvanderw@hsph.harvard.edu

**James M. Robins**
Department of Epidemiology
Department of Biostatistics
Harvard School of Public Health
robins@hsph.harvard.edu

## Abstract

Identifying effects of actions (treatments) on outcome variables from observational data and causal assumptions is a fundamental problem in causal inference. This identification is made difficult by the presence of confounders which can be related to both treatment and outcome variables. Confounders are often handled, both in theory and in practice, by adjusting for covariates, in other words considering outcomes conditioned on treatment and covariate values, weighed by probability of observing those covariate values. In this paper, we give a complete graphical criterion for covariate adjustment, which we term the adjustment criterion, and derive some interesting corollaries of the completeness of this criterion.

## 1 Introduction

Estimating causal effects in the presence of confounding variables is an important problem in causal inference. By causal effect here we mean a probability distribution on some set of outcomes after administering treatment. Confounders are found among those variables which are related to both treatment and outcome. Pearl [7], [8] developed a system of inference about such problems using the language of causal diagrams. Though complete rules for identification of causal effects have been derived using this language [5], [13], [12], [14] in practice the vast majority of causal effect identification is done either via instrumental variable methods, or by covariate adjustment. Pearl's back-door criterion [7] gives graphical conditions for when covariate adjustment yields valid functionals for the causal effect. The back-door criterion, however, is not complete ([8], chapter 11). In other words, there exist causal diagrams where the back-door criterion fails for some set of covariates, yet adjusting for that set results in a valid functional for the causal effect.

In this paper, we give a complete graphical criterion for the validity of covariate adjustment as an identification method for causal effects, which we term the adjustment criterion. We further give some interesting corollaries of the completeness of our criterion.

## 2 Preliminaries

Causal diagrams [7], [8] are directed acyclic graphs representing probabilistic causal domains. Nodes in causal diagrams represent variables of interest while arrows represent direct causal influences. Causal diagrams are associated with structural causal models (SCMs) defined in [8], chapter 7. SCMs consist of a set of observed variables $\mathbf{V}$ and unobserved or latent variables $\{U_1, ..., U_k\} = \mathbf{U}$. Without loss of generality, it is usually assumed that the unobserved variables are exogenous and drawn from a distribution $P(\mathbf{u}) = \prod_i P(u_i)$, and the observed variables are endogenous, whose values are obtained from the values of other variables by means of determining functions. The distribution $P(\mathbf{u})$ over the unobserved variables, together with the functions onto the endogenous variables, define an observed distribution $P(\mathbf{v})$. Exogenous and endogenous variables together can be thought of as drawn from a joint distribution $P(\mathbf{u}, \mathbf{v})$. When we mention models or causal models in this paper, we will be referring to structural causal models.

Each causal model induces a causal diagram, as follows. Each variable in the model is represented by a vertex, and a vertex corresponding to a variable $V_i$ has arrows incoming from every variable whose value is used to determine the value of $V_i$ by its determining function. In such a graph exogenous variables have no incoming arrows. In this paper we will restrict our attention to acyclic causal diagrams.

We will also use standard graph-theoretic terminology.

A path is a sequence of edges where each pair of adjacent edges in the sequence share a node, and each such shared node can only occur once in the path. A path from a node $X$ to a node $Y$ consists exclusively of directed arrows pointing away from $X$ is called directed or causal, all other kinds of paths are called non-causal. An ancestor of $X$ is any node which has a directed path to $X$ (including $X$ itself). A descendant of $X$ is any node which $X$ has a directed path to (including $X$ itself). The ancestor set of a node set $\mathbf{X}$ is all ancestors of any node in $\mathbf{X}$. The descendant set of a node set $\mathbf{X}$ is all descendants of any node in $\mathbf{X}$. We will abbreviate ancestor and descendant sets of $\mathbf{X}$ in $G$ as $An(\mathbf{X})_G$ and $De(\mathbf{X})_G$.

One of the advantages of the causal diagram representation is its ability to display conditional independence among variables in terms of a path separation criterion known as d-separation [6]. Two sets of nodes $\mathbf{X}, \mathbf{Y}$ are said to be d-separated by a third set $\mathbf{Z}$ if every path from any node $X \in \mathbf{X}$ to any node $Y \in \mathbf{Y}$ is "blocked". A path is "blocked" if it contains a triple of consecutive nodes connected in one of the following three ways: $W_i \to W_j \to W_k$, $W_i \leftarrow W_j \to W_k$, and $W_i \to W_l \leftarrow W_k$, where $W_j \in \mathbf{Z}$, and neither $W_l$ nor any descendant of $W_l$ is in $\mathbf{Z}$. Paths (or node sets) that are not d-separated are called d-connected, active, or open. A path starting with an outgoing arrow is called a front-door path, while a d-connected path starting with an incoming arrow is called a back-door path. The relationship between d-separation and conditional independence is provided by the following well-known theorem [6].

**Theorem 1 (Pearl)** *Let $G$ be a causal diagram. Then in any model $M$ with a distribution $P(\boldsymbol{u}, \boldsymbol{v})$ inducing $G$, if $\boldsymbol{X}$ is d-separated from $\boldsymbol{Y}$ by $\boldsymbol{Z}$ in $G$, then $\boldsymbol{X}$ is independent of $\boldsymbol{Y}$ given $\boldsymbol{Z}$ in $P(\boldsymbol{u}, \boldsymbol{v})$.*

Following [3], we will denote conditional independence statements like the one in the theorem by $(\mathbf{X} \perp\!\!\!\perp \mathbf{Y} | \mathbf{Z})$. A graph $G$ which represents independences of $P(\mathbf{v})$ in $M$ via d-separation is called an *I-map* of $P(\mathbf{v})$ [6].

If many of the variables in a causal diagram are latent, it can be inconvenient to consider long path stretches containing nothing but latent variables. One graphical representation that avoids this is the *latent projection* [19]. A latent projection of a causal diagram is a mixed graph containing directed and bidirected arcs, where there is a vertex for every observable node, and two observable nodes $X, Y$ are connected by a directed arrow if there is a d-connected path from $X$ to $Y$ in the original causal diagram containing only arrows pointing away from $X$ and towards $Y$, and each node on this path other than $X$ and $Y$ is latent. Similarly, $X$ and $Y$ are connected by a bidirected arrow if there is a d-connected path from $X$ to $Y$ in the original causal diagram which starts with an arrow pointing to $X$ and ends in an arrow pointing to $Y$, where each node on this path other than $X$ and $Y$ is latent. D-separation generalizes in a straightforward way to latent projections [9], and latent projections preserve all d-separation statements of the original causal diagrams. In the remainder of the paper we will assume all graphs are latent projections.

Consider graphs, shown in Fig. 1 (a), (b) representing the effect of a drug ($X$) on an outcome ($Y$), in the presence of a chronic exposure ($Z$), say long term alcohol abuse. We assume the temporal order of chronic exposure and drug exposure is not known. If the chronic exposure precedes the drug, as in Fig. 1 (a), it affects both the drug reaction (via liver damage) and the outcome. If the chronic exposure follows the drug, as in Fig. 1 (b), it does not affect the outcome, since the outcome is measured soon after drug exposure (not giving liver damage from alcohol time to accumulate). These two models disagree on observational claims which means they can be tested (for instance by checking if $Z$ is independent of $Y$ given $X$). However, assume further that we lack the power to distinguish these models, yet wish to make inferences about the causal effect of $X$ on $Y$.

We can analyze this problem using the $do(\mathbf{x})$ notation, introduced by Pearl [8] (and equivalent to $g(\mathbf{x})$ notation in [10]). In this notation, $do(\mathbf{x})$ stands for an idealized experiment, or *intervention*, where the values of $\mathbf{X}$ are set to $\mathbf{x}$, regardless of the normal causal mechanisms which determine the values of $\mathbf{X}$. Pearl formalized the *causal effect* of $\mathbf{X}$ on $\mathbf{Y}$ as the distribution of $\mathbf{Y}$ after the $do(\mathbf{x})$ operation, written $P(\mathbf{y}|do(\mathbf{x}))$. In general, $P(\mathbf{y}|do(\mathbf{x}))$ is different from the conditional distribution of $\mathbf{Y}$ given $\mathbf{X}$, or $P(\mathbf{y}|\mathbf{x})$, since $\mathbf{X}$ may be related to $\mathbf{Y}$ both via causal action and via certain variables which affect both $\mathbf{X}$ and $\mathbf{Y}$. In identifying $P(\mathbf{y}|do(\mathbf{x}))$ we are interested only in effects of $X$ on $Y$ along causal paths.

A classic problem in causal inference is identifying $P(\mathbf{y}|do(\mathbf{x}))$ from observational data, since implementing idealized experiments in practice is difficult. A well-known way of identifying $P(\mathbf{y}|do(\mathbf{x}))$ in situations where $\mathbf{Y}$ and $\mathbf{X}$ share common causes is performing covariate adjustment, in other words finding a set of variables $\mathbf{Z}$ such that $P(\mathbf{y}|do(\mathbf{x})) = \sum_{\mathbf{z}} P(\mathbf{y}|\mathbf{x}, \mathbf{z}) P(\mathbf{z})$. Pearl developed the well-known back-door criterion [7], [8], which gave graphical conditions for when the above adjustment formula yields valid functionals for $P(\mathbf{y}|do(\mathbf{x}))$, where $\mathbf{Y}, \mathbf{X}$ are variable sets.

**Definition 1 (back-door criterion)** *Let $\boldsymbol{X}, \boldsymbol{Y}$ be disjoint sets of nodes in a graph $G$. Then a set $\boldsymbol{Z}$*

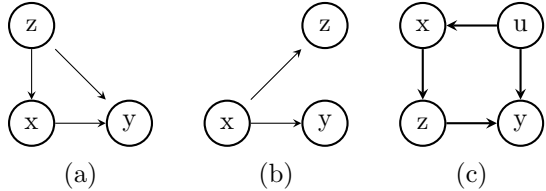

Figure 1: (a),(b) Causal diagrams where the effect $P(y|do(x))$ can be identified via $\sum_z P(y|x,z)P(z)$. (c) A causal diagram $G$ where $P(y|do(x))$ is not identifiable via adjustment for any set.

*satisfies the back-door criterion relative to* $(\boldsymbol{X}, \boldsymbol{Y})$ *if*

- *No element in $\boldsymbol{Z}$ is a descendant of $\boldsymbol{X}$*
- *$\boldsymbol{Z}$ d-separates all back-door path from $\boldsymbol{X}$ to $\boldsymbol{Y}$*

**Theorem 2 (Pearl)** *If some set $\boldsymbol{Z}$ satisfies the back-door criterion relative to $(\boldsymbol{X}, \boldsymbol{Y})$ in the graph $G$, then $P(\boldsymbol{y}|do(\boldsymbol{x})) = \sum_{\boldsymbol{z}} P(\boldsymbol{y}|\boldsymbol{x},\boldsymbol{z})P(\boldsymbol{z})$ in every model which induces $G$.*

The back-door criterion allows us to conclude that $P(y|do(x)) = \sum_z P(y|x,z)P(z)$ in Fig. 1 (a). The back-door criterion, however, is not complete. In particular, $P(y|do(x)) = \sum_z P(y|x,z)P(z)$ in Fig. 1 (b), though $Z$ does not satisfy the back-door criterion with respect to $(X, Y)$.

In fact, the process of adjustment itself is not complete in that there are certain graphs where a causal effect $P(\mathbf{y}|do(\mathbf{x}))$ is identifiable, but there is no $\mathbf{Z}$ for which $\sum_{\mathbf{z}} P(\mathbf{y}|\mathbf{x},\mathbf{z})P(\mathbf{z})$ is a valid functional for $P(\mathbf{y}|do(\mathbf{x}))$. This is true of a graph shown in Fig. 1 (c), where we are interested in $P(y|do(x))$ and $U$ is unobserved (which means we cannot adjust for it). In fact, the valid functional for this causal effect is $\sum_z P(z|x) \sum_{x'} P(y|z,x')P(x')$, which is obtained from the front-door criterion [8].

In general, there are complete algorithms which, given a graph and a causal effect of interest, produce a functional for the causal effect valid given the causal assumptions embodied by the graph [18], [13],[12]. By complete, we mean that any time the algorithm fails, there is no valid functional for the effect of interest in all models with the given graph, and thus no other algorithm could produce such a functional. Despite these results, in practical causal inference problems identification is generally done via covariate adjustment. This is because in practice it is very difficult to determine precise causal relationships between variables to produce a graph, while it is much easier to talk about "coarser" causal notions such as pre-treatment covariates and confounding.

For this reason, we concentrate on covariate adjustment in this paper. After introducing relevant terminology in subsequent sections, we will develop a criterion which will precisely characterize when covariate adjustment for $\mathbf{Z}$ yields correct functionals for $P(\mathbf{y}|do(\mathbf{x}))$ in $G$.

## 3 Graphs and Counterfactuals

In this section, we will introduce counterfactual distributions, which generalize causal effects, and counterfactual graphs, which display independences in counterfactual distributions.

As mentioned in the previous section, the responses of the remaining observable variables other than $\mathbf{X}$ to the intervention are represented by an interventional distribution denoted as $P(\mathbf{v}\backslash\mathbf{x}|do(\mathbf{x}))$ or $P_{\mathbf{x}}(\mathbf{v}\backslash\mathbf{x})$. The response of a single observable variable $Y$ to $do(\mathbf{x})$ is sometimes denoted with a counterfactual variable $Y_{\mathbf{x}}$. Similarly, the response of a set $\mathbf{Y} = \{Y^1, Y^2, ..., Y^k\}$ to $do(\mathbf{x})$ will be denoted by $\mathbf{Y}_{\mathbf{x}}$.

The $do(\mathbf{x})$ operation in a graph $G$ can be represented graphically by removing all arrows pointing to nodes in $\mathbf{X}$. This is because ordinarily, values of $\mathbf{X}$ are determined by direct causal influences stemming from its parent variables in the graph, and these same influences are ignored by the $do(\mathbf{x})$ operation. We will denote the resulting graph, which is sometimes called the mutilated graph, as $G_{\overline{\mathbf{x}}}$.

Once we fix the value of every exogenous variable in the model, the remaining variables become deterministically fixed. This allows us to use the distribution $P$ over exogenous variables to define joint distributions over counterfactual variables, even if the interventions which determine these variables disagree with each other. Such a joint distribution is defined as follows:

$$P(Y^1_{\mathbf{x}^1} = y^1, ..., Y^k_{\mathbf{x}^k} = y^k) = \sum_{\{\mathbf{u}|Y^1_{\mathbf{x}^1}(\mathbf{u}) = y^1 \wedge ... \wedge Y^k_{\mathbf{x}^k}(\mathbf{u}) = y^k\}} P(\mathbf{u})$$

where $\mathbf{U}$ is the set of exogenous variables in the model.

Independence among counterfactual variables can be displayed by a special graph, called the *counterfactual graph* [15], [14], just as independence between causal variables is displayed by a causal diagram via d-separation. The construction of the counterfactual graph is rather intricate, however for the purposes of this paper it is sufficient to consider a simplified version known as the *twin network graph* [2], [1].

The twin network graph displays counterfactual independence among two possible worlds, the pre-intervention world which is represented by the original graph $G$, and the post-intervention world, which is represented by the graph $G_{\overline{\mathbf{x}}}$. In other words, the twin network graph is an I-map for the joint counterfactual distribution $P(\mathbf{v}, \mathbf{v_x})$, where $\mathbf{V}$ is the set of all observable variables, and $\mathbf{V_x}$ is the set of all observable variable after the intervention $do(\mathbf{x})$ was performed. The observable nodes in these two graphs share the $\mathbf{U}$ variables, to signify common history of these worlds up to the point divergence due to $do(\mathbf{x})$. We will add an additional refinement, appearing in [15], where node copies of all non-descendants of $\mathbf{X}$ in $G$ are merged in the twin network graph (since such nodes are the same random variable in either intervened or natural world). Fig. 2 shows the twin network graph for $do(x)$ in the graph shown in Fig. 1 (a) (note that $Z$ is shared across both worlds).

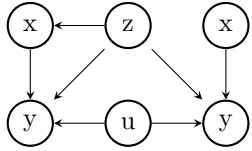

Figure 2: The twin network graph for $do(x)$ in the graph in Fig. 1 (a).

We are introducing the mathematics of counterfactuals in this paper because, in the course of proving the soundness of our criterion for $\mathbf{Z}$ serving as a valid adjustment set for treatments $\mathbf{X}$ and outcomes $\mathbf{Y}$, we will show that if the criterion is used in structural causal models, it will imply that $(\mathbf{Y_x} \perp\!\!\!\perp \mathbf{X}|\mathbf{Z})$, in other words the post intervention outcomes are independent of treatments conditioned on the covariates. This assumption is sometimes known as *conditional ignorability* [11], and is often assumed in practice as a prelude for using adjustment functionals to identify causal effects. This implication will show that conditional ignorability is the logically minimal assumption which permits adjustment in structural causal models, in other words if it's true, adjustment is valid, and valid adjustment implies its truth.

## 4 The Adjustment Criterion

Before we can give a complete graphical criterion, we need to introduce some helpful terminology.

**Definition 2 (collider)** *A three-node path $(X, Y, Z)$ from $X$ to $Z$ is called a collider if the edges connecting $X$ to $Y$ and $Y$ to $Z$ have arrowheads pointing to $Y$.*

**Definition 3 (inducing path)** *A path $\pi$ from node set $\mathbf{X}$ to node set $\mathbf{Y}$ in a causal graph $G$ is called inducing if:*

- *Every three-node path segment of $\pi$ is a collider*
- *Every node on the path is ancestral to $\mathbf{X}$ or $\mathbf{Y}$*

**Definition 4 (proper causal path)** *Let $\mathbf{X}, \mathbf{Y}$ be sets of nodes. A causal path from a node in $\mathbf{X}$ to a node in $\mathbf{Y}$ is called proper if it does not intersect $\mathbf{X}$ except at the end point.*

A complete adjustment criterion would, for any effect $P(\mathbf{y}|do(\mathbf{x}))$, allow any adjustment set $\mathbf{Z}$ which opens all proper causal paths, but blocks "everything else."

**Definition 5 (adjustment criterion)** *$\mathbf{Z}$ satisfies the adjustment criterion relative to $(\mathbf{X}, \mathbf{Y})$ in $G$ if*

- *No element in $\mathbf{Z}$ is a descendant in $G_{\overline{\mathbf{x}}}$ of any $W \notin \mathbf{X}$ which lies on a proper causal path from $\mathbf{X}$ to $\mathbf{Y}$.*
- *All non-causal paths in $G$ from $\mathbf{X}$ to $\mathbf{Y}$ are blocked by $\mathbf{Z}$.*

It's fairly clear from the definition that this criterion generalizes the back-door criterion. In particular, note that while all back-door paths are non-causal, not all non-causal paths are back-door. For example, in the Fig. 1 (c), there is one causal path from $X$ to $Z$, namely $X \to Z$, and one non-causal path from $Z$ to $X$, namely $Z \to Y \leftarrow U \to X$. This path is not back-door with respect to $Z$ since the edge closest to $Z$ points away from $Z$. In Fig. 1 (b), $Z$ satisfies the adjustment criterion with respect to $(X, Y)$, yet does not satisfy the back-door criterion with respect to $(X, Y)$ (since $Z$ is a descendant of $X$).

**Lemma 1** *If $\mathbf{Z}$ satisfies the back-door criterion with respect to $(\mathbf{X}, \mathbf{Y})$ in $G$, then $\mathbf{Z}$ satisfies the adjustment criterion with respect to $(\mathbf{X}, \mathbf{Y})$ in $G$.*

*Proof:* This follows by above definition and definition of d-separation. □

In the next section, we show that not only does the adjustment criterion always yield valid functionals for the causal effect, but whenever it fails on some covariate set in a graph, there is some model which induces that graph where adjustment by the covariate set will not yield the correct causal effect expression.

## 5 Soundness and Completeness

We first prove the completeness direction.

**Theorem 3** *Assume the adjustment criterion fails to hold for $\mathbf{Z}$ with respect to $(\mathbf{X}, \mathbf{Y})$ in $G$. Then there exists a model inducing $G$ such that $P(\mathbf{y}|do(\mathbf{x})) \neq \sum_{\mathbf{z}} P(\mathbf{y}|\mathbf{x},\mathbf{z})P(\mathbf{z})$.*

*Proof:* We have two cases to consider. Either $\mathbf{Z}$ contains a descendant of a node on a proper causal path from $\mathbf{X}$ to $\mathbf{Y}$, and its closest ancestor on this causal path is not in $\mathbf{X}$, or no element of $\mathbf{Z}$ is a descendant of a node on a proper causal path from $X \in \mathbf{X}$ to $Y \in \mathbf{Y}$ such that its closest ancestor on this causal path is not $X$, but $\mathbf{Z}$ opens a non-causal path from $\mathbf{X}$ to $\mathbf{Y}$.

For the first case, fix a graph containing only the proper causal path from $X \in \mathbf{X}$ to $Y \in \mathbf{Y}$ with appropriate segment connecting this path to $Z \in \mathbf{Z}$. See Fig. 3 for an example. Without loss of generality we can assume a single $Y$ in this graph. We get $P(y|do(x)) = P(y|x)$.

It's not difficult to find a model with this graph where $\sum_z P(y|z,x)P(z)$ to not equal $P(y|x)$. For instance let the node on the path from $X$ to $Y$ that is the closest to $Z$ be denoted $W$. If $W = Z$, then $P(y|z,x) = P(y|z)$, so $\sum_z P(y|z,x)P(z) = P(y)$, which is easy to arrange to not equal $P(y|x)$. Otherwise, $P(y|x) = \sum_w P(y|w)P(w|x)$, while $\sum_z P(y|x,z)P(z) = \sum_{z,w} P(y,x,z,w)/P(x|z) = \sum_w P(y|w) \sum_z P(w|x,z)P(z)$.

We can view $\sum_w P(y|w)$ as an operator which maps probability distributions over $W$ onto probability distributions over $Y$. It is not difficult to arrange $P(y|w)$ in such a way that this map is one-to-one. Then all we have to show is that $P(w|x) = \sum_z P(w|x,z)P(z|x)$ need not equal $\sum_z P(w|x,z)P(z)$ in every model. But this is also not hard to show – let $\sum_z P(w|x,z)$ be one-to-one and make sure $P(z|x)$ is not equal to $P(z)$.

For the second case, assume $\mathbf{Z}$ opens a non-causal path $\pi$ from $X \in \mathbf{X}$ to $Y \in \mathbf{Y}$. Consider the subgraph $G'$ of $G$ containing all nodes in $\pi$, along with a path for every open collider in $\pi$, witnessing this collider as ancestral of $\mathbf{Z}$. Since the first case does not hold, this subgraph cannot contain nodes on any proper causal path from $X$ to $Y$ (otherwise, either $\pi$ contains a proper causal path as a subpath, or $\mathbf{Z}$ contains descendants of nodes on the proper causal path and the closest ancestor on the proper causal path is not $X$). See Fig. 4 for an example. Thus, in $G'$, $P(y|do(x)) = P(y)$. Let the subset of $\mathbf{Z}$ intersecting $G'$ be denoted $\mathbf{Z}'$.

It is a simple matter to pick the behavior of the variables on the open path from $X$ to $Y$ such that $P(y|x,\mathbf{z}')$ is not equal to $P(y|\mathbf{z}')$, and the mappings $\sum_{\mathbf{z}'} P(y|x,\mathbf{z}')$ and $\sum_{\mathbf{z}'} P(y|\mathbf{z}')$ do not map a given $P(\mathbf{z})$ to the same output. This implies that $\sum_{\mathbf{z}'} P(y|x,\mathbf{z}')P(\mathbf{z}')$ is not equal to $P(y)$ in the model.

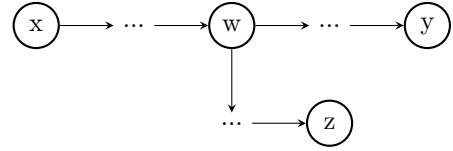

Figure 3: Case 1 of theorem 3.

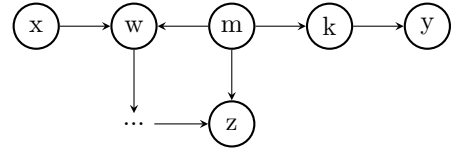

Figure 4: Case 2 of theorem 3.

So far, we have shown that if the adjustment criterion for $\mathbf{Z}$ with respect to $(\mathbf{X}, \mathbf{Y})$ is violated in $G$, we can find a path from $X \in \mathbf{X}$ to $Y \in \mathbf{Y}$ such that there is a model inducing a subgraph $G'$ associated with this path where the causal effect $P(y|do(x))$ is not equal to the adjustment functional $\sum_{\mathbf{z}} P(y|x,\mathbf{z})P(\mathbf{z})$. But to prove the general completeness statement, we need only find *any* model where adjustment fails to equal the causal effect expression. In particular, we are free to pick models not faithful [17] to $G$. In fact, assume all nodes in $G$ outside of $G'$ are mutually independent.

Then for case one, we have the following identities: $\sum_{\mathbf{z}} P(\mathbf{y}|\mathbf{x},\mathbf{z})P(\mathbf{z}) = \sum_z P(y|z,x)P(z)P(\mathbf{y} \setminus \{y\})$, and $P(\mathbf{y}|do(\mathbf{x})) = P(y|x)P(\mathbf{y} \setminus \{y\})$. For case two, we have: $\sum_{\mathbf{z}} P(\mathbf{y}|\mathbf{x},\mathbf{z})P(\mathbf{z}) = \sum_{\mathbf{z}'} P(y|x,\mathbf{z}')P(\mathbf{z}')P(\mathbf{y} \setminus \{y\})$, and $P(\mathbf{y}|do(\mathbf{x})) = P(y)P(\mathbf{y} \setminus \{y\})$. Our result then follows for both cases in such a model. □

Before showing soundness, we give some definitions and utility lemmas.

A path is a sequence of edges where each pair of adjacent edges in the sequence share a node, and each such shared node can only occur once in the path.

**Definition 6 (route)** *A route from $X$ to $Y$ in a graph $G$ is a sequence of edges in $G$, where each pair of adjacent edges share a node, the unshared node of the first edge is $X$, and the unshared node of the last edge is $Y$.*

*Note: shared nodes can occur more than once.*

A route is d-separated if the same triples are blocked as in the definition of d-separated paths on p. 2. The difference between a *route* and a *path* is that path sequences cannot contain duplicate nodes while route sequences can.

**Definition 7 (direct route)** *Let $\pi$ be a route from $X$ to $Y$ in $G$. Label each node occurrence in the route*

$\pi$ by the number of times the node has already occurred earlier in $\pi$ (so the first occurrence is labeled 0, the second is labeled 1, etc.)

A direct route $\pi^*$ is a subsequence obtained from $\pi$ inductively as follows:

- First node in $\pi^*$ is the first node in $\pi$ with the largest occurrence number.

- If the $k$th shared node in $\pi^*$ (and $m$th node in $\pi$) is $(X_i, r)$, and $X_i \neq Y$, let the $k+1$th node in $\pi^*$ be $(X_j, n)$, where $X_j$ is the $m+1$th node in $\pi$, and $n$ is the largest occurrence number of $X_j$ in $\pi$.

**Lemma 2** *For every route $\pi$ in $G$, the direct route $\pi^*$ is a path. Moreover, if $\pi$ is d-connected, then $\pi^*$ is d-connected.*

*Proof:* That the direct route is a route follows from the previous definition – direct routes have node neighbors only if those nodes were neighbors in the original route, and those nodes are neighbors in $G$. The direct route is a path since the definition only takes nodes in the original sequence with the largest occurrence numbers.

To show the second part, we must show all three-node sequences in $\pi^*$ are open. Consider all elements of $\mathbf{Z}$ on the route $\pi$. Since $\pi$ is open, all edges connecting nodes in $\pi$ neighboring these elements in $\mathbf{Z}$ must point to them. Since every edge in $\pi^*$ is inherited from $\pi$, this implies that if $\mathbf{Z}$ is a part of $\pi^*$, it must be in the middle of a collider.

What remains to show is that a collider in $\pi^*$ which does not have an element of $\mathbf{Z}$ as a middle node, must contain such a node as a descendant. Consider the two edges $e_1, e_2$ making up such a collider in the original route $\pi$. If these edges are adjacent in $\pi$, we are done, since $\pi$ is active. Otherwise, let $e_3$ be an edge following $e_1$ in $\pi$, and $e_e$ an edge preceding $e_2$ in $\pi$. If either $e_1, e_3$ or $e_4, e_2$ form a collider we are also done. Otherwise, either $e_3$ or $e_4$ must point to an ancestor of $\mathbf{Z}$, if $\pi$ is active. □

We are now ready to show soundness.

**Theorem 4** *Assume the adjustment criterion holds for $\mathbf{Z}$ and $(\mathbf{X}, \mathbf{Y})$ in $G$. Then for every model inducing $G$, $(\mathbf{Y_x} \perp\!\!\!\perp \mathbf{X}|\mathbf{Z})$.*

*Proof:* Assume the adjustment criterion holds in $G$. Consider the twin network graph $G^*$ showing the pre-intervention world, and the world after $do(\mathbf{x})$. We want to show that if $\mathbf{Z}$ satisfies the adjustment criterion for $(\mathbf{X}, \mathbf{Y})$ in $G$, then $\mathbf{X}$ is d-separated from $\mathbf{Y_x}$ by $\mathbf{Z}$ in the twin network graph. We will show the contrapositive, namely that an active path from $\mathbf{X}$ to $\mathbf{Y_x}$ in $G^*$ implies the criterion is violated.

Fix an active path $\pi$ from $\mathbf{X}$ to $\mathbf{Y_x}$ in $G^*$. Without loss of generality, assume $\pi$ intersects $\mathbf{X}$ only once at the endpoint. Since we are conditioning on variables $\mathbf{Z}$ in the pre-intervention world, no descendant of $\mathbf{X}$ in the post-intervention world is conditioned on. Thus, any active path $\pi$ from $\mathbf{X}$ to $Y_\mathbf{x}$ that "lands" in the post-intervention world must descend, via arrows pointing to $Y_\mathbf{x}$, to $Y_\mathbf{x}$.

By above, an active path $\pi$ from $X$ to $Y_\mathbf{x}$ in $G^*$ consists of three parts, $\pi_1, \pi_2, \pi_3$, where $\pi_1$ is just some active path in the original graph $G$, $\pi_3$ is a directed path in $G_{\overline{\mathbf{x}}}$ where every node in $\pi_3$ is a descendant of $\mathbf{X}$, and $\pi_2$ is a single edge connecting these paths in $G^*$ to form $\pi$.

Note that $G^*$ contains copies of nodes in $G$, with one copy corresponding to the variable in the original world, and another copy corresponding to the variable in the post-intervention world. Thus, even though $\pi$ is a path in $G^*$, $\pi$ may contain counterfactual nodes (say $W_\mathbf{x}$ and $W$), which refer to the same node in $G$. Let $\pi'$ be a sequence of node/occurrence number pairs in $G$ obtained from $\pi$ by the following two steps. The first step replaces all occurrences of counterfactual variables in $\pi$ by the nodes in $G$ these variables were derived from, coupled with appropriate occurrence numbers. The second step replaces all occurrences of a single variable twice in a row in the sequence obtained from step one by a single occurrence of that node (with occurrence numbers of this node and all subsequent occurrences appropriately decremented). It's not difficult to see that $\pi'$ is a route in $G$. Let $\pi^*$ be the direct route of $\pi'$.

Note that the sections of $\pi'$ corresponding to sections $\pi_1$ and $\pi_3$ in $\pi$ must be active. The remaining section in $\pi'$ is a node triple containing a middle node whose copy was pointed to by $\pi_2$ in $\pi$. If this node triple exists in $\pi$, it must also be active in $\pi'$. Otherwise, it was obtained from four consecutive nodes in $\pi$, with two middle ones being node copies connected by $\pi_2$. These middle nodes cannot intersect or be ancestral of $\mathbf{Z}$ (or the adjustment criterion is violated). Otherwise, the triple in $\pi'$ must look like $\circ \leftarrow \circ \rightarrow \circ$, which implies it's active. Thus, $\pi'$ is an active route.

By Lemma 2, $\pi^*$ is an active path. If $\pi^*$ is non-causal, we are done, as this means there is an active non-causal path connecting $\mathbf{Y}$ and $\mathbf{X}$ given $\mathbf{Z}$, which contradicts the adjustment criterion. If $\pi^*$ is causal, we can assume without loss of generality it is also proper causal. We now have two cases: either $\pi_2$ is bidirected or directed. Since $\pi^*$ is causal, the first edge in $\pi$ must be an arrow leading away from an element in $\mathbf{X}$. But if $\pi_2$ is bidirected, then the only way $\pi$ could be active in $G^*$ is if the second node in $\pi$ is an ancestor of $\mathbf{Z}$.

But by construction of $\pi^*$, the second node in $\pi$ is also the second node in $\pi^*$ (although perhaps with a different occurrence number). But this contradicts the adjustment criterion.

If the $\pi_2$ edge is a directed arc, then the nodes it connects are a parent/child pair in $G$, with the parent copy (say of node $M$) in the $G$ part of $G^*$, and the child copy (say of node $N$) in the $G_{\overline{x}}$ part of $G^*$. Then $M$ cannot have $\mathbf{X}$ ancestors, otherwise it would not be in the $G$ section of $G^*$. If $\pi_1$ and $\pi_3$ do not share nodes, then $\pi^*$ cannot be a proper causal path from $\mathbf{X}$ to $\mathbf{Y}$ since otherwise every node in $\pi^*$, including $M$, must be a descendant of $\mathbf{X}$, which is a contradiction.

If $\pi_1$ and $\pi_3$ share nodes, then the only way to reach $M$ from $\mathbf{X}$ with an active path that starts with an arrow pointing away from $\mathbf{X}$ is via a collider open by $\mathbf{Z}$. But this implies the second node in $\pi$ (and thus the second node in $\pi^*$) is an ancestor of $\mathbf{Z}$, which contradicts the adjustment criterion. □

**Corollary 1** *Assume the adjustment criterion holds for $\mathbf{Z}$ and $(\mathbf{X}, \mathbf{Y})$ in $G$. Then for every model inducing $G$, $\sum_{\mathbf{z}} P(\mathbf{y}|\mathbf{x}, \mathbf{z})P(\mathbf{z})$.*

*Proof:* The assumption implies $(\mathbf{Y_x} \perp\!\!\!\perp \mathbf{X}|\mathbf{Z})$ in every model inducing $G$, by Theorem 4. By laws of probability, $P(\mathbf{Y_x}) = \sum_{\mathbf{z}} P(\mathbf{Y_x}|\mathbf{z})P(\mathbf{z})$. The axiom of consistency (called composition in [8], chapter 7) then implies the conclusion. □

## 6 Corollaries

We now derive some interesting corollaries of the completeness of the adjustment criterion. First, we show the logical minimality of the conditional ignorability assumption for covariate adjustment.

**Corollary 2** $P(\mathbf{y}|do(\mathbf{x})) = \sum_{\mathbf{z}} P(\mathbf{y}|\mathbf{x},\mathbf{z})P(\mathbf{z})$ *in every model inducing $G$, if and only if $(\mathbf{Y_x} \perp\!\!\!\perp \mathbf{X}|\mathbf{Z})$ holds in every model inducing $G$,*

*Proof:* If $(\mathbf{Y_x} \perp\!\!\!\perp \mathbf{X}|\mathbf{Z})$ holds in every model, the adjustment formula is valid by proof of Corollary 1. If the adjustment formula is valid for every model inducing $G$, then the adjustment criterion holds in $G$ by Theorem 3, and the adjustment formula holds for every model inducing $G$ by Corollary 1. □

This corollary is useful in practical causal inference situations where we are unwilling to commit to a particular causal diagram due to lack of domain knowledge, yet feel confident in making limited counterfactual independence assumptions. This corollary states that if we want to identify effects via covariate adjustment in such situations, we might as well assume conditional ignorability, since any other assumption we may use that permits adjustment will imply it.

Next, we consider the problem of what adjustment sets are sufficient for an unbiased estimate of causal effect, given that we know that a valid adjustment set exists. We give some utility lemmas first.

**Lemma 3** *A node set $\mathbf{X}$ cannot be d-separated from a node set $\mathbf{Y}$ in $G$ if and only if there exists an inducing path from $\mathbf{X}$ to $\mathbf{Y}$ in $G$.*

*Proof:* See proof of Theorem 10 in [16]. □

**Lemma 4** *If $\mathbf{X}$ and $\mathbf{Y}$ are not connected by an inducing path in $G$ then $\mathbf{A} = An(\mathbf{X} \cup \mathbf{Y}) \setminus (\mathbf{X} \cup \mathbf{Y})$ is a set sufficient for d-separating $\mathbf{X}$ and $\mathbf{Y}$.*

*Proof:* Any path from $\mathbf{X}$ to $\mathbf{Y}$ which involves any nodes not in $\mathbf{A}$ must contain a collider and so isn't d-connecting. Consider a path from $\mathbf{X}$ to $\mathbf{Y}$ consisting entirely of nodes in $\mathbf{A}$ (we can assume the path only intersects $\mathbf{X}$ and $\mathbf{Y}$ at the endpoints). Since we condition on $\mathbf{A}$, the d-connecting path must contain only colliders, but this implies there is a path consisting entirely of colliders all of which are ancestral of $\mathbf{X}$ or $\mathbf{Y}$. This contradicts the absence of an inducing path. □

**Lemma 5** *Let $G$ be a graph, $P(\mathbf{y}|do(\mathbf{x}))$ a causal effect. Let $\mathbf{M}$ be the set of nodes on the proper causal paths from $\mathbf{X}$ to $\mathbf{Y}$. Let $G^*$ be the latent projection obtained from $G$ by marginalizing out $\mathbf{M}$.*

*Then $\mathbf{Z}$ satisfies the adjustment criterion with respect to $(\mathbf{X}, \mathbf{Y})$ in $G$ if and only if $\mathbf{Z}$ also satisfies the adjustment criterion with respect to $(\mathbf{X}, \mathbf{Y})$ in $G^*$.*

*Proof:* Note that $\mathbf{Z}$ cannot intersect $\mathbf{M}$ or the adjustment criterion is violated. Since marginalization does not affect d-separation statements, the second part of the criterion is preserved. Similarly, marginalization does not create new ancestor/descendant relations that were not present before marginalization. Thus the first part of the criterion is preserved as well. □

The following result suggests that the back-door criterion, despite being incomplete, does capture the intuition behind valid adjustment in some sense, since whenever a valid adjustment set exists for a causal effect, the subset of this adjustment set which forms the non-descendants of treatment variables is also a valid adjustment set.

**Theorem 5** *Fix sets $\mathbf{X}, \mathbf{Y}$ in $G$ such that $P(\mathbf{y}|do(\mathbf{x})) = \sum_{\mathbf{z}} P(\mathbf{y}|\mathbf{x},\mathbf{z})P(\mathbf{z})$ for some set $\mathbf{Z}$ in every causal model inducing $G$. Let $\mathbf{Z}^* = \mathbf{Z} \setminus De(\mathbf{X})$. Then $\mathbf{Z}^*$ satisfies the back-door criterion with respect to $(\mathbf{X}, \mathbf{Y})$.*

*Proof:* Consider some d-connected path $\pi$ from $\mathbf{X}$ to $\mathbf{Y}$, given $\mathbf{Z}^*$. If $\pi$ starts with an arrow pointing away from $\mathbf{X}$ then, since $\mathbf{Z}^*$ contains no descendants of $\mathbf{X}$, $\pi$ is either proper causal to $\mathbf{Y}$, or non-causal and blocked.

If $\pi$ starts with an arrow pointing towards $\mathbf{X}$, it is a back-door path. Since $\mathbf{Z}^*$ contains no descendants of $\mathbf{X}$, and $\mathbf{Z}^*$ is the only set we are conditioning on, once $\pi$ reaches a node $W$ which is a descendant of $\mathbf{X}$, in order for it to stay active, it must descend along directed arrows away from $W$ to $\mathbf{Y}$. But no element of $\mathbf{Z} \setminus \mathbf{Z}^*$ can block the portion of $\pi$ from $W$ to $\mathbf{Y}$ because that portion of the path is on a proper causal path from $\mathbf{X}$ to $\mathbf{Y}$ (since $W$ is a descendant of $\mathbf{X}$). Thus $\pi$ must have been active conditioned on $\mathbf{Z}$. But by assumption $\mathbf{Z}$ satisfies the adjustment criterion for $(\mathbf{X}, \mathbf{Y})$, and we have a contradiction. □

**Theorem 6** *Let $G$ be a graph, and $P(\boldsymbol{y}|do(\boldsymbol{x}))$ a causal effect. Assume a set of nodes $\mathbf{Z}$ exists such that $P(\boldsymbol{y}|do(\boldsymbol{x})) = \sum_{\boldsymbol{z}} P(\boldsymbol{y}|\boldsymbol{x}, \boldsymbol{z}) P(\boldsymbol{z})$ in every model inducing $G$. Let $\mathbf{M}$ be the set of nodes which lie on some proper causal path from $\mathbf{X}$ to $\mathbf{Y}$ in $G$. Then $P(\boldsymbol{y}|do(\boldsymbol{x})) = \sum_{\boldsymbol{a}} P(\boldsymbol{y}|\boldsymbol{x}, \boldsymbol{a}) P(\boldsymbol{a})$ in every model inducing $G$, where $\mathbf{A} = An(\mathbf{X} \cup \mathbf{Y}) \setminus (\mathbf{X} \cup \mathbf{Y} \cup \mathbf{M})$.*

*Proof:* By Theorem 3, $\mathbf{Z}$ satisfies the adjustment criterion with respect to $(\mathbf{X}, \mathbf{Y})$ in $G$. Let $G^*$ be the graph obtained from $G$ by marginalizing out $\mathbf{M}$. By Lemma 5, $\mathbf{Z}$ still satisfies the adjustment criterion with respect to $(\mathbf{X}, \mathbf{Y})$ in $G^*$. By Theorem 5, $\mathbf{Z}^* = \mathbf{Z} \setminus De(\mathbf{X})$ satisfies the back-door criterion with respect to $(\mathbf{X}, \mathbf{Y})$.

Let $G^*_{\underline{\mathbf{x}}}$ be the graph obtained from $G^*$ by cutting all outgoing directed arrows from $\mathbf{X}$. By construction, $\mathbf{Z}$ d-separates $\mathbf{X}$ and $\mathbf{Y}$ in $G^*_{\underline{\mathbf{x}}}$. This implies by Lemma 3 that $\mathbf{X}$ and $\mathbf{Y}$ have no inducing path in $G^*_{\underline{\mathbf{x}}}$. This, in turn, implies $An(\mathbf{X} \cup \mathbf{Y}) \setminus (\mathbf{X} \cup \mathbf{Y})$ must also d-separate $\mathbf{X}$ from $\mathbf{Y}$ in $G^*_{\underline{\mathbf{x}}}$, by Lemma 4. But this set is precisely equal to $\mathbf{A}$, and by construction of $G^*_{\underline{\mathbf{x}}}$, it's clear $\mathbf{A}$ satisfies the back-door criterion with respect to $(\mathbf{X}, \mathbf{Y})$ in $G^*$. Finally, Lemmas 1 and 5 tell us $\mathbf{A}$ satisfies the adjustment criterion with respect to $(\mathbf{X}, \mathbf{Y})$ in $G$. This proves our result. □

Theorem 6 says that if we know that a valid adjustment set exists, we are free to adjust for all ancestors of treatment and outcome, as long as those ancestors do not occupy proper causal paths we are interested in. Adjusting for all observable causal ancestors $\mathbf{A}$ can be statistically desirable when constructing estimators for causal effects $P(\mathbf{y}|do(\mathbf{x}))$, since such a set often includes many independent causes of outcome nodes $\mathbf{Y}$, and including such nodes in the adjustment set can improve the efficiency of the estimator. Furthermore, Theorem 6 can be useful in practical causal inference problems, where it is often assumed that a valid co-variate set exists, and it is known which covariates are ancestral of treatments and outcomes, while the exact causal relationships between covariates themselves are not known.

## 7 A Non-counterfactual Soundness Proof

In the previous section, we assumed structural causal models, which imply counterfactual quantities, and we heavily used these quantities in our proofs. This allowed us to derive logical minimality of the conditional ignorability assumption for covariate adjustment (Corollary 2). Nevertheless, the use of counterfactuals has come under some criticism in the literature [4], due to their untestability. There exist alternative formulations of causal models [8], [17], [4] which either do not make use of counterfactuals at all. It is therefore of interest to examine whether our results are applicable to such models without making use of counterfactuals. This should be possible since soundness and completeness of the adjustment criterion are facts about interventional and observational distributions, not specifically about counterfactuals.

In this section, we give such an alternative proof.

**Definition 8 (magnification)** *Let $G$ be a causal diagram, let $\mathbf{E} = \{e_1, ..., e_k\}$ be a set of edges in $G$. Let $G_{\boldsymbol{e}}$ be a directed acyclic graph constructed as follows. $G_{\boldsymbol{e}}$ contains every observable node in $G$, a new observable node $W_e$ for every bidirected arc $e$, and possibly a new observable node $C_i$ for every directed arc $e_i \in \mathbf{E}$. Moreover, in $G_{\boldsymbol{e}}$, if an edge is not in $\mathbf{E}$, it exists in $G_{\boldsymbol{e}}$. For every bidirected arc $e$ in $G$ connecting $X$ and $Y$, there is a directed edge in $G_{\boldsymbol{e}}$ from $W_e$ to both $X$ and $Y$, and for every $e_i \in \mathbf{E}$, connecting $X$ and $Y$ in $G$, there is a directed arc in $G_{\boldsymbol{e}}$ from $X$ to $C_i$ and from $C_i$ to $Y$.*

*Then $G_{\boldsymbol{e}}$ is the magnification of $G$ with respect to $\mathbf{E}$.*

A magnification of a graph is a new graph where we "unmarginalize" all unobserved variables, and moreover introduce a new mediator for every previously direct causal arrow from $X$ to $Y$ in $\mathbf{E}$. We call this graph a magnification because by using this new graph we are increasing the granularity at which we are looking at our causal model represented by $G$.

**Theorem 7** *Let $G$ be a causal diagram, and $P(\boldsymbol{y}|do(\boldsymbol{x}))$ a causal effect of interest, and $\mathbf{Z}$ a set partitioned into non-descendants of $\mathbf{X}$ ($\mathbf{Z}_{nd}$), and descendants of $\mathbf{X}$ ($\mathbf{Z}_d$). Let $\mathbf{E}$ be the set of all directed edges pointing away from $\mathbf{Y}$. Let $G_{\boldsymbol{e}}$ be the magnification of $G$ with respect to $\mathbf{E}$. Assume there exists a set of nodes $\mathbf{L}$ such that*

(a) $\mathbf{L} \cup \mathbf{Z}_{nd}$ satisfy the back-door criterion with respect to $(\mathbf{X}, \mathbf{Y})$

(b) $(\mathbf{X} \perp\!\!\!\perp \mathbf{L} | \mathbf{Z})$

(c) $(\mathbf{Y} \perp\!\!\!\perp \mathbf{Z}_d | \mathbf{X}, \mathbf{Z}_{nd}, \mathbf{L})$

where statements using the $\perp\!\!\!\perp$ symbol in (b) and (c) are to be interpreted as d-separation statements in $G_e$. Then $P(\mathbf{y}|do(\mathbf{x})) = \sum_{\mathbf{z}} P(\mathbf{y}|\mathbf{x}, \mathbf{z}) P(\mathbf{z})$ in any causal model $M$ inducing $G$.

*Proof:* Fix some model $M_\mathbf{e}$ which induces $G_\mathbf{e}$ which recovers the marginal $P(\mathbf{v})$ of $M$ when marginalizing over all newly introduced nodes in $G_\mathbf{e}$ over $G$. In this model, we have: $P(\mathbf{y}|do(\mathbf{x})) = \sum_{\mathbf{l}, \mathbf{z}_{nd}} P(\mathbf{y}|\mathbf{x}, \mathbf{l}, \mathbf{z}_{nd}) P(\mathbf{l}, \mathbf{z}_{nd})$, by (a). $P(\mathbf{y}|do(\mathbf{x})) = \sum_{\mathbf{l}, \mathbf{z}} P(\mathbf{y}|\mathbf{x}, \mathbf{l}, \mathbf{z}) P(\mathbf{l}, \mathbf{z})$, by (c). Finally, $P(\mathbf{y}|do(\mathbf{x})) = \sum_{\mathbf{l}, \mathbf{z}} P(\mathbf{y}|\mathbf{x}, \mathbf{l}, \mathbf{z}) P(\mathbf{l}|\mathbf{z}, \mathbf{x}) P(\mathbf{z}) = \sum_{\mathbf{z}} P(\mathbf{y}|\mathbf{x}, \mathbf{z}) P(\mathbf{z})$, by (b). Since adjustment holds in $M_\mathbf{e}$, and only makes use of variables in $M$, adjustment also holds in $M$. □

What we will show next is the adjustment criterion in $G$ implies we can always find such an $\mathbf{L}$ in $G_\mathbf{e}$. This means that if we cannot find such an $\mathbf{L}$, the adjustment criterion fails in $G$, and it's completeness (the proof of which did not use counterfactuals) will imply the completeness of this alternative criterion. We prove the implication without using counterfactuals.

**Theorem 8** *Assume the adjustment criterion holds for $\mathbf{Z}$ with respect to $(\mathbf{X}, \mathbf{Y})$ in $G$. Then we can find $\mathbf{L}$ satisfying the properties in Theorem 7 in $G_e$.*

*Proof:* Let $\mathbf{L}$ be the intersection of the following three sets: ancestors of $\mathbf{Z}$ in $G_\mathbf{e}$, non-descendants of $\mathbf{X}$ in $G_\mathbf{e}$, and nodes d-connected to $\mathbf{Y}$ given $\mathbf{Z}$ in the graph $G_\mathbf{e} \setminus \mathbf{X}$ where $\mathbf{X}$ and all edges pointing to $\mathbf{X}$ are removed. We will show this set satisfies (a), (b) and (c).

(a) implies that $\mathbf{L} \cup \mathbf{Z}_{nd}$ consists of non-descendants of $\mathbf{X}$ only. This is true by definition for both sets. (a) also implies that $\mathbf{L} \cup \mathbf{Z}_{nd}$ block all back-door paths from $\mathbf{X}$ to $\mathbf{Y}$. Assume this isn't true, and fix an open back-door path from $\mathbf{X}$ to $\mathbf{Y}$ given $\mathbf{L}, \mathbf{Z}_{nd}$. We know from Theorem 5 that $\mathbf{Z}_{nd}$ is back-door for $(\mathbf{X}, \mathbf{Y})$. That means $\mathbf{Z}_{nd}$ by itself blocks this path, and $\mathbf{L}$ is what opens it. So the only way nodes in $\mathbf{L}$ are involved in the back-door path are by being descendants of colliders on this path. $\mathbf{L}$ is an ancestor of either $\mathbf{Z}$ not through $\mathbf{X}$, or of $\mathbf{X}$. In the former case, $\mathbf{Z}$ opens the collider on the path. In the latter case, we restart the argument with a back-door path from $X \in \mathbf{X}$ between $\mathbf{L}$ and $\mathbf{Z}$. If all colliders are thus open, we get a contradiction since we assumed all non-causal paths, and in particular all back-door paths from $\mathbf{X}$ to $\mathbf{Y}$ are blocked by $\mathbf{Z}$.

Assume (b) isn't true. Then there is an open path from $\mathbf{X}$ to $\mathbf{L}$ that is either causal, or non-causal. Since $\mathbf{L}$ is not a descendant of $\mathbf{X}$, this path is not causal. Otherwise, there is a non-causal path from $\mathbf{X}$ to $L \in \mathbf{L}$ (given $\mathbf{Z}$), and an open path from $L \in \mathbf{L}$ to $\mathbf{Y}$ (given $\mathbf{Z}$). Adjoining these two d-connected paths creates a d-connected route $\pi$ from $\mathbf{X}$ to $\mathbf{Y}$. The direct route $\pi^*$ of $\pi$ is then a d-connected path given $\mathbf{Z}$ from $\mathbf{X}$ to $\mathbf{Y}$ by Lemma 2. Moreover, since the path from $L$ to $\mathbf{Y}$ is in $G \setminus \mathbf{X}$, the direct route is non-causal, and we have a contradiction.

Assume (c) isn't true. Then there is an open path from $\mathbf{Z}_d$ to $\mathbf{Y}$ that is either causal or non-causal. If it's causal, then the adjustment criterion is violated, since $\mathbf{Z}_d$ will block a proper causal path from $\mathbf{X}$ to $\mathbf{Y}$. If the path $\pi$ from $\mathbf{Z}_d$ to $\mathbf{Y}$ is non-causal, then it's most general form is: $Z_d o \to Z_1 \leftarrow ... \to Z_k \leftarrow oY$. Here circles at the edge ends of the first and last edge of the path mean that end of edge can be either an arrow or a tail, and elements $Z_1, ... Z_k$ are in $\mathbf{X} \cup \mathbf{Z}_{nd} \cup \mathbf{L}$.

Consider the $Z_i \in \mathbf{X}$ closest (on $\pi$) to $Y$ among all $Z_i$ which belong to $\mathbf{X}$. Then we restart our argument considering a back-door path from this $Z_i$ to $Y$. Thus we can assume $Z_1, ..., Z_k$ does not intersect $\mathbf{X}$. Similarly, if $Z_i \in \mathbf{L}$, $Z_i$ must be ancestral of $\mathbf{Z}$. If it's only ancestral via an element in $\mathbf{X}$, we consider the $Z_i$ closest to $Y$ along $\pi$ with this property, and conclude there is a non-causal path from that $X$ to $Y$ that's open given $\mathbf{Z}$. This contradicts the adjustment criterion we assumed. If $Z_i$ is ancestral of $\mathbf{Z}$ not via $\mathbf{X}$, then if this $Z$ is in $\mathbf{Z}_d$, we repeat the argument with this $Z_d$ being the new endpoint of a non-causal path to $Y$. Otherwise, $Z$ is in $\mathbf{Z}_{nd}$. Thus, without loss of generality assume all $\mathbf{Z}_1, ..., \mathbf{Z}_k$ are in $\mathbf{Z}_{nd}$.

We have two cases, the path ends in $\to Y$, or $\leftarrow Y$. In the former case, if the highest point on the $Z_k \leftarrow ... \to Y$ segment is a descendant of $X$, then $Z_k$ cannot be in $\mathbf{Z}_{nd}$ as was our assumption. If the highest point is not a descendant of $X$, it is in $\mathbf{L}$ by definition (since it is d-connected to Y given $\mathbf{Z}$ in $G \setminus \mathbf{X}$, is an ancestor of $\mathbf{Z}$, and not a descendant of $\mathbf{X}$), and so is conditioned on and blocks this path. If the path ends in $\leftarrow Y$, then if this $Y$ is a descendant of $\mathbf{X}$, the adjustment criterion is violated since this $\mathbf{Y}$ must be an ancestor of $\mathbf{Z}$. If $Y$ is not a descendant of $\mathbf{X}$, then it's mediator variable we added to $G_\mathbf{e}$ which connects $Y$ to its second node on the path cannot be a descendant of $X$ either, and so is in $\mathbf{L}$ and blocks the path. □

## 8 Discussion

We have given a complete criterion for covariate adjustment in structural causal models. This criterion generalizes the well-known back-door criterion [7], to certain cases where its permissible to adjust for descendants of treatment variables. Our criterion implies that whenever we adjust using a (valid) covariate set which includes descendants of the treatment variable, it is possible to remove these descendants from the covariate set, without affecting validity of adjustment. Additionally, we show that the existence of a valid adjustment set of any kind implies adjusting for all ancestors of treatment and outcome that do not lie on causal paths of interest is sufficient to give a valid estimate of the causal effect. We give both a counterfactual proof of our result, which implies the logical minimality of the conditional ignorability assumption for covariate adjustment, and a non-counterfactual proof which holds in models which remain agnostic about counterfactual statements.

We believe characterizing covariate adjustment is important because, despite the fact that more sophisticated identification methods exist [18], [12], the vast majority of causal effect identification in practice is done via covariate adjustment.

Our result allows the use of covariate adjustment even if the model is partially unspecified or misspecified, in other words if the direction of some arrows in the causal diagram is unknown or incorrect. Going back to our example in Fig. 1, our results allow us to use the adjustment formula safely, even if we are not sure whether the graph in Fig. 1 (a) or the graph in Fig. 1 (b) is the correct model, and lack the statistical power to distinguish these cases.

Finally, since our result can be used to list all possible valid covariate sets for a given causal effect, it paves the way for considering which set yields the most statistically desirable estimator of a causal effect, for instance in terms of efficiency, or mean squared error.